\documentclass[prd,aps,preprintnumbers,preprint,floatfix,superscriptaddress]{revtex4}

\usepackage{amsmath,amssymb}
\usepackage{cancel}

\begin{document}

\title{Bulk Fermions in Soft Wall Models}

\author{S. Mert Aybat}
\affiliation{Institute for Theoretical Physics, ETH, CH-8093,
  Z\"urich, Switzerland}
\author{Jos\'e Santiago}
\affiliation{Institute for Theoretical Physics, ETH, CH-8093,
  Z\"urich, Switzerland}
\affiliation{
CAFPE and Departamento de F\'{\i}sica Te\'orica y del 
Cosmos, \\
Universidad de Granada, E-18071 Granada, Spain
}

\preprint{CAFPE-126/09}
\preprint{UGFT-256/09}

\begin{abstract}
We discuss the implementation of bulk fermions in soft wall
models. The introduction of a position dependent bulk mass allows for
a well defined Kaluza-Klein expansion for bulk fermions. The
realization of flavor and the contribution to electroweak precision
observables are shown to be very similar to the hard wall case. The
bounds from electroweak precision test are however milder with gauge
boson Kaluza-Klein modes as light as $\sim 1.5$ TeV compatible with current
experimental bounds.
\end{abstract}

\maketitle


Models with warped extra dimensions
offer a new approach to solve the
hierarchy problem, explaining the stability of the electroweak scale
against the ultraviolet physics simply due to the geometry of
space-time in these models. They also have a very appealing flavor
structure, as they naturally predict hierarchical fermion masses 
with an extra built-in flavor protection for light fermions. 
Hard wall models~\cite{Randall:1999ee}, based on a slice of AdS$_5$,
are compatible with current constraints and a natural realization of
flavor for  
masses of the gauge boson Kaluza-Klein (KK) modes heavier than 
$m_{KK}\ge 3.5$ TeV~\cite{Carena:2006bn} (see~\cite{Davoudiasl:2009cd}
for a recent review). 
Soft wall models, which modify infrared
(IR) physics by removing the IR brane and changing the gravitational
background, have been shown to be compatible with current data for
much lighter gauge boson KK modes $m_{KK}\ge 0.5$
TeV~\cite{Falkowski:2008fz} when fermions are constrained to live on the
UV brane. In this talk we discuss how to incorporate bulk fermions in
the picture and the impact they have on constraint on the model from
electroweak precision tests (EWPT) and flavor physics. More details can be
found in~\cite{MertAybat:2009mk} (see also~\cite{Delgado:2009xb} for other
approaches to bulk fermions in soft wall models).

The soft wall is realized on an AdS$_5$ background
\begin{equation}
ds^2=a^2(z)\left[dx^2-dz^2\right]\,,
\end{equation}
where the warp factor is $a(z)=L_0/z$ and $L_0$ corresponds to the
inverse curvature scale of the AdS$_5$ space (and gives the location
of the UV brane on the extra dimension $L_0\leq z \leq \infty$). The
departure from pure AdS is given by a
dilaton with profile $\Phi=\left(z/L_1\right)^2$, so that the matter
action reads 
\begin{equation}
S_{matter}=\int d^5x \sqrt{g}{\rm e}^{-\Phi}\mathcal{L}_{matter}\,.
\end{equation}
$L_1$ is the scale at which the background begins to depart from pure
AdS. 
The expected values of these scales for a natural theory of
electroweak symmetry breaking (EWSB) are $L_0^{-1}\sim
M_{\mathrm{Pl}}$ and $L_1^{-1} \sim $ TeV.

Let us now consider bulk fermions in this background. The action of a
bulk fermion, $\Psi(x,z)$, reads
\begin{equation}
S=
\int d^5x \, a^4 e^{-\Phi} \bar{\Psi} \left[
i \cancel{\partial} +\left( \partial_5 + 2 \frac{a^\prime}{a} - \frac{1}{2}
\Phi^\prime \right) \gamma^5 - a M \right] \Psi 
=\int d^5x \, \bar{\psi} \big[
i \cancel{\partial} + \partial_5 \gamma^5 - a M \big] \psi, 
\label{bulk:fermion:action}
\end{equation}
where in the second equation we have defined
\begin{equation}
\psi(x,z)\equiv a^{2}(z) e^{-\Phi(z)/2} \Psi(x,z).
\end{equation}
In order to obtain a sensible KK expansion of this bulk fermion with
the required features, we postulate a $z$-dependent bulk Dirac mass of
the form
\begin{equation}
 M(z)=\frac{c_0}{L_0} + \frac{c_1}{L_0} \frac{z^2}{L_1^2}\,,
\end{equation}
where $c_{0,1}$ are dimensionless constants expected to be order
one. The equations of motion derived from the fermionic action read 
\begin{equation}
i\cancel{\partial} \psi_{L,R} +(\pm \partial_5-aM)\psi_{R,L}=0,
\end{equation}
where $ \psi_{L,R} \equiv \frac{1\mp \gamma^5}{2} \psi$.
A standard expansion in KK modes,
\begin{equation}
\psi_{L,R}(x,z) = \sum_n f_n^{L,R}(z) \psi^{(n)}_{L,R}(x),
\end{equation}
with $i \cancel{\partial} \psi^{(n)}_{L,R}(x) = m_n
\psi^{(n)}_{R,L}(x)$ gives the equations for the fermionic profiles
\begin{equation}
(\partial_5 \pm a M)f_n^{L,R} = \pm m_n f_n^{R,L}.
\end{equation}
The orthonormality condition
\begin{equation}
\int_{L_0}^\infty f_n^L f_m^L = \int_{L_0}^\infty f_n^R f_m^R = 
\delta_{nm},
\end{equation}
then gives the action as a sum over four-dimensional Dirac KK modes and
 possibly massless zero modes,
\begin{equation}
S=\int d^4x \, \sum_n \bar{\psi}^{(n)} [ i \cancel{\partial} - m_n ]
\psi^{(n)}.
\end{equation}
The first order coupled equations for the fermionic profiles can 
be iterated to give two decoupled second order differential equations 
\begin{equation}
\Big[\partial_5^2 \pm (a M)^\prime -(aM)^2 +m_n^2\Big] f_n^{L,R}(z) =0.
\end{equation}
Inserting the expression of the metric and the mass, we get for the LH
profile,
\begin{equation}
\left[
\partial_5^2 - \frac{c_0(c_0+1)}{z^2} + \frac{c_1}{L_1^2} (1-2c_0) +
m_n^2 -\frac{c_1^2 z^2}{L_1^4} \right] f_n^L = 0\,,
\end{equation}
while the RH solution is identical to the LH one with the identification
$c_{0,1} \to -c_{0,1}$.
The normalizable solutions of the coupled linear equations can then be
written as,
\begin{eqnarray}
&& \left . 
\begin{array}{l}
f_n^L(z)=
N_n z^{-c_0} e^{-\frac{c_1z^2}{2L_1^2}} 
U\left( -\frac{L_1^2 m_n^2}{4 c_1},\frac{1}{2}-c_0,\frac{c_1z^2}{L_1^2}
\right),
\\
f_n^R(z)=
N_n \frac{m_n}{2} z^{1-c_0} e^{-\frac{c_1z^2}{2L_1^2}} 
U\left( 1-\frac{L_1^2 m_n^2}{4 c_1},\frac{3}{2}-c_0,\frac{c_1z^2}{L_1^2}
\right),
\end{array}
\right\} \Rightarrow \mbox{ for } c_1 > 0, \\ 
&& \left . 
\begin{array}{l}
f_n^L(z)=
-N_n \frac{m_n}{2} z^{1+c_0} e^{\frac{c_1z^2}{2L_1^2}} 
U\left( 1+\frac{L_1^2 m_n^2}{4 c_1},\frac{3}{2}+c_0,-\frac{c_1z^2}{L_1^2}
\right),
\\
f_n^R(z)=
N_n z^{c_0} e^{\frac{c_1z^2}{2L_1^2}} 
U\left( \frac{L_1^2 m_n^2}{4 c_1},\frac{1}{2}+c_0,-\frac{c_1z^2}{L_1^2}
\right),
\end{array}
\right\} \Rightarrow \mbox{ for } c_1 < 0, 
\end{eqnarray}
where $U(a,b,z)$ is the confluent hypergeometric function and
the normalization constants $N_n$ are fixed by normalizing either
the LH or the RH profile. 

The masses and the possible presence of zero modes is determined
by the boundary conditions (bc). 
It is easy to see (details can be found in~\cite{MertAybat:2009mk}) 
the qualitative
and quantitative equivalence of bc
\begin{equation}
[\pm,\pm]_{\mathrm{hard wall}} \Leftrightarrow 
[\pm,\mathrm{sign}(c_1)]_{\mathrm{soft wall}}.
\end{equation}
Here $[\pm,\pm]$ denote the bc at the UV and IR brane, respectively,
where a $+$ ($-$) means that the RH (LH) chirality has
Dirichlet bc (it vanishes) at the corresponding brane. 
For instance, in the hard wall we have a LH (RH) chiral zero mode for
$[++]$ ($[--]$) bc. Similaly, in the soft wall we find the following
zero modes
\begin{equation}
f_0^{L,R}(c_0,c_1;z)=\left[\frac{L_0^{1\mp 2c_0}}{2} E_{\pm c_0+\frac{1}{2}}
  \left(\pm c_1\frac{L_0^2}{L_1^2} \right) \right]^{-\frac{1}{2}}
z^{\mp c_0}
e^{\mp\frac{c_1 z^2}{2L_1^2}},
\end{equation}
where 
$E_\nu(z)= \int_1^\infty dt\, e^{-zt}/t^\nu$ is the Exponential
Integral E function.
A LH zero mode exists if $c_1>0$ and the UV bc is $[+]$,
whereas a RH zero mode exists if $c_1<0$ and the UV bc is $[-]$, just
as in the hard wall.
Once the right boundary conditions for the existence of a chiral zero
mode are imposed, we see that $c_1$ controls the exponential die-off
in the IR whereas $c_0$ controls the localization of the zero mode.
The same equivalence also occurs at the level of massive modes.
 
This KK expansion can be now used to study the phenomenology of bulk
fermions in soft wall models. The most important features from the
phenomenological point of view are the masses and couplings of the
bulk fermions to the Higgs and gauge bosons. The masses scale
according to Regge trajectories, as governed by the soft wall
\begin{equation}
m_n \sim \frac{\sqrt{n}}{L_1}.
\end{equation}
This means that the spectrum of new massive fermions is more packed,
with more modes accessible at colliders and more modes giving indirect
contributions to EWPT at loop level. Indeed, a detailed calculation of
the two most relevant observables, the $T$ parameter and the
$Zb_L\bar{b}_L$ coupling, using the results in~\cite{Anastasiou:2009rv}
shows that up to $\sim 10$ modes can contribute before the results
stabilize~\cite{MertAybat:2009mk}.
Regarding the couplings, both gauge and Yukawa couplings share the
main features that make hard wall models so appealing. Gauge couplings
to gauge boson KK modes are almost universal for UV localized fermions
with departures from universality proportional to the fermion
masses. This guarantees a flavor protection mechanism similar to the
one that makes hard wall models compatible with flavor data and a low
scale of new physics with minimal tuning or structure. Similarly,
Yukawa couplings naturally predict hierarchical masses and mixing
angles so again the realization of flavor is as natural in soft wall
models as it is with a hard wall.

We have performed a detailed fit to all relevant electroweak precision
observables, including the one loop contribution from the top sector
to the $T$ parameter and the $Zb_L\bar{b}_L$ coupling.  We performed the one 
loop calculations for these observables treating the EWSB perturbatively. We also checked 
the validity of this approximation, and even by including a small number of KK modes
the perturbative treatment of EWSB is a good approximation up to per mille level. 
The result is
that minimal models with a custodial protection of these two
observables~\cite{Agashe:2003zs,Agashe:2006at} are compatible with
EWPT provided the gauge boson KK modes are heavier
than
\begin{equation}
m_n^{\mathrm{GB}} \gtrsim 1.5-3~\mathrm{TeV},
\end{equation}
depending on the details of the Higgs
boson~\cite{MertAybat:2009mk}. Less minimal models or different
backgrounds may  allow even lighter KK modes. 
This milder constraint and the fact that masses scale with $\sqrt{n}$
makes the LHC prospects of discovering these models very exciting.

\begin{acknowledgments}

This work has been partially supported by SNSF under contract
200021-117873, MICINN under the RyC program
and project FPA2006-05294 and Junta de Andaluc\'{\i}a projects FQM
101, FQM 437 and FQM03048. 
\end{acknowledgments}

\end{document}